
\magnification 1200

\font\eightrm=cmr8
\font\eighti=cmmi8
\font\eightsy=cmsy8
\font\eightbf=cmbx8
\font\eighttt=cmtt8
\font\eightit=cmti8
\font\eightsl=cmsl8
\font\sixrm=cmr6
\font\sixi=cmmi6
\font\sixsy=cmsy6
\font\sixbf=cmbx6
\catcode`@11
\newskip\ttglue

\def\eightpoint{\def\rm{\fam0\eightrm}
\textfont0=\eightrm \scriptfont0=\sixrm \scriptscriptfont0=\fiverm
\textfont1=\eighti \scriptfont1=\sixi \scriptscriptfont1=\fivei
\textfont2=\eightsy \scriptfont2=\sixsy \scriptscriptfont2=\fivesy
\textfont3=\tenex \scriptfont3=\tenex \scriptscriptfont3=\tenex
\textfont\itfam=\eightit \def\it{\fam\itfam\eightit}
\textfont\slfam=\eightsl \def\sl{\fam\slfam\eightsl}
\textfont\ttfam=\eighttt \def\tt{\fam\ttfam\eighttt}
\textfont\bffam=\eightbf
\scriptfont\bffam=\sixbf
\scriptscriptfont\bffam=\fivebf \def\bf{\fam\bffam\eightbf}
\tt \ttglue=.5em plus.25em minus.15em
\normalbaselineskip=6pt
\setbox\strutbox=\hbox{\vrule height7pt width0pt depth2pt}
\let\sc=\sixrm \let\big=\eightbig \normalbaselines\rm}
\newinsert\footins
\def\newfoot#1{\let\@sf\empty
  \ifhmode\edef\@sf{\spacefactor\the\spacefactor}\fi
  #1\@sf\vfootnote{#1}}
\def\vfootnote#1{\insert\footins\bgroup\eightpoint
  \interlinepenalty\interfootnotelinepenalty
  \splittopskip\ht\strutbox 
  \splitmaxdepth\dp\strutbox \floatingpenalty\@MM
  \leftskip\z@skip \rightskip\z@skip
  \textindent{#1}\footstrut\futurelet\next\fo@t}
\def\fo@t{\ifcat\bgroup\noexpand\next \let\next\f@@t
  \else\let\next\f@t\fi \next}
\def\f@@t{\bgroup\aftergroup\@foot\let\next}
\def\f@t#1{#1\@foot}
\def\@foot{\strut\egroup}
\def\footstrut{\vbox to\splittopskip{}}
\skip\footins=\bigskipamount 
\count\footins=1000 
\dimen\footins=8in 

\def\ref#1{$^{#1}$}
\def\flex{\raise 6pt\hbox{$\leftrightarrow $}\! \! \! \! \! \! }
\def\oversome#1{ \raise 8pt\hbox{$\scriptscriptstyle #1$}\! \! \! \! \! \! }
\def\tr{ \mathop{\rm tr}}

\newbox\bigstrutbox
\setbox\bigstrutbox=\hbox{\vrule height10pt depth5pt width0pt}
\def\bigstrut{\relax\ifmmode\copy\bigstrutbox\else\unhcopy\bigstrutbox\fi}
\def\refer[#1/#2]{ \item{#1} {{#2}} }
\def\rev<#1/#2/#3/#4>{{\it #1\/} {\bf#2}, {#3}({#4})}
\def\boxit#1{\vbox{\hrule\hbox{\vrule\kern3pt
\vbox{\kern3pt#1\kern3pt}\kern3pt\vrule}\hrule}}

\def\2figure#1#2#3#4{\vbox{ \hrule width#1truecm \hbox{\vrule height#2truecm
\hskip #1truecm
\vrule height#2truecm }\hrule width#1truecm \hbox{\vrule\vbox{\hsize #1truecm
\baselineskip=10pt
\noindent\strut#3}\vrule}\hrule width#1truecm
\hbox{\vrule\vbox{\hsize #1truecm
\baselineskip=10pt
\noindent\strut#4}\vrule}\hrule width#1truecm  }}
\def\3figure#1#2#3#4#5{\vbox{ \hrule width#1truecm \hbox{\vrule height#2truecm
\hskip #1truecm
\vrule height#2truecm }\hrule width#1truecm \hbox{\vrule\vbox{\hsize #1truecm
\baselineskip=10pt
\noindent\strut#3}\vrule}\hrule width#1truecm
 \hbox{\vrule\vbox{\hsize #1truecm
\baselineskip=10pt
\noindent\strut#4}\vrule}
\hrule width#1truecm \hbox{\vrule\vbox{\hsize #1truecm
\baselineskip=10pt
\noindent\strut#5}\vrule}\hrule width#1truecm  }}

\def\sqr#1#2{{\vcenter{\hrule height.#2pt
   \hbox{\vrule width.#2pt height#1pt \kern#1pt
    \vrule width.#2pt}
    \hrule height.#2pt}}}


\font\sf=cmss10

\def\smin{\,\raise 0.06em \hbox{${\scriptstyle \in}$}\,}
\def\smsubset{\,\raise 0.06em \hbox{${\scriptstyle \subset}$}\,}

\def\Natural{\hbox{\hskip 1.5pt\hbox to 0pt{\hskip -2pt I\hss}N}}
\def\Integer{\>\hbox{{\sf Z}} \hskip -0.82em \hbox{{\sf Z}}\,}
\def\Rational{\hbox{\hbox to 0pt{\hskip 2.7pt \vrule height 6.5pt
                                  depth -0.2pt width 0.8pt \hss}Q}}
\def\Real{\hbox{\hskip 1.5pt\hbox to 0pt{\hskip -2pt I\hss}R}}
\def\Complex{\hbox{\hbox to 0pt{\hskip 2.7pt \vrule height 6.5pt
                                  depth -0.2pt width 0.8pt \hss}C}}

\def \E {{{\rm e}}}

\def \shg {\sqrt {\hat g } }


\def \1ok{{1\over \kappa ^2} }

\def \3dslim {{\rm DS}\!\!\!\!\!\!\!\!\lim }
\def \4dslim {{\rm DS}\!\!\!\!\!\!\!\!\!\!\lim }
\def \tr {{\rm tr}\, }

\def \2kk{\left( \matrix {2k\cr k\cr }\right) }
\def \Rs4{{R^k\over 4^k} }

\def \1ok{{1\over \kappa ^2} }



\hsize 6truein
\vsize 8.5truein
\baselineskip 13.3 truept
\def \E {{{\rm e}}}
\hfill CERN-TH.7161/94
\nopagenumbers
\centerline{\bf NON-CRITICAL SUPERSTRINGS: A COMPARISON BETWEEN}

\centerline {\bf CONTINUUM AND DISCRETE APPROACHES}
\vskip 1.5truecm

\centerline{A. ZADRA}
\vskip .3truecm

\centerline{Instituto de F\'\i sica, University of S\~ao Paulo}

\centerline{CP 20516, S\~ao Paulo, Brazil}

\vskip 1truecm

\centerline{E. ABDALLA\newfoot{$^\dagger $}{Permanent address: Instituto de
F\'\i sica, University of S.Paulo. Work supported by Fapesp.}}
\vskip .3truecm

\centerline {CERN-TH, CH-1211 Geneva 23, Switzerland}

\vskip 2truecm

\centerline {\bf Abstract}
\vskip 1truecm

\noindent
We review the relation between the matrix model and Liouville approaches to
two-dimensional gravity as elaborated by Moore, Seiberg and Staudacher. Then,
based on the supersymmetric Liouville formulation and the discrete eigenvalue
model proposed by Alvarez-Gaum\'e, Itoyama, Ma\~nes and Zadra, we extend the
previous relation to the supersymmetric case. The minisuperspace approximation
for the supersymmetric case is formulated, and the corresponding wave equation
is found.
\vskip 6truecm
\noindent CERN-TH.7161/94

\noindent February 1994

\vfill \eject
\countdef\pageno=0 \pageno=1
\newtoks\footline \footline={\hss\tenrm\folio\hss}
\def\folio{\ifnum\pageno<0 \romannumeral-\pageno \else\number\pageno \fi}
\def\advancepageno{\ifnum\pageno<0 \global\advance\pageno by -1
\else\global\advance\pageno by 1 \fi}

\noindent 1. {\bf Introduction}
\vskip .3truecm

\noindent It is generally accepted that the $m$-th multicritical point of the
Hermitian 1-matrix model corresponds to the coupling of the Liouville theory to
perturbations of the $(2,2m-1)$ minimal conformal model. However it took some
time to understand this relation. A precise dictionary between Liouville and
matrix model operators, correlation functions and loop amplitudes was proposed
by Moore, Seiberg and Staudacher in Ref.  [1]. They recognized how part of the
difficulties were related to contact terms. After a careful examination of
macroscopic loop amplitudes in two-dimensional gravity, which must satisfy the
Wheeler-de Witt (WdW) equation, they managed to determine a new frame of
scaling operators and couplings (called {\it conformal frame}) in the matrix
theory, whose correlation functions and scaling dimensions were in perfect
agreement with the Liouville predictions.

This successful program motivated us to generalize their computations for the
supersymmetric case. We considered the amplitudes of the N = 1 super-Liouville
theory coupled to $c<3/2$ matter, which were calculated in Ref.  [2]. As for
the
discrete counterpart, there is no actual generalized matrix model. So far
the most successful description has been given by the discrete model introduced
by Alvarez-Gaum\'e, Itoyama, Ma\~nes and Zadra in Ref. [3]
in terms of super-eigenvalue variables. Nevertheless we shall prove that, in
spite of the missing  interpretation in terms of matrices, this
discrete theory also shows a complete agreement with continuum results.

We dedicate sect. 2 to a brief review of the purely bosonic case, emphasizing
the main results from the continuum approach in subsect. 2.1 and describing in
subsect. 2.2 the discrete theory and the procedure that determines the
conformal
frame of operators. In sect. 3 we present the supersymmetric version of the
models: the super Liouville theory is summarized in subsect. 3.1; the
super-eigenvalue model, its standard scaling operators and their correlation
functions are reviewed in subsect. 3.2; then, in subsect. 3.3 we follow the
strategies  of the bosonic case to define a superconformal frame; their wave
functions are studied in subsect. 3.4. We leave sect. 4 for the final comments
and a brief conclusion.

\vskip .5truecm
\noindent 2. {\bf Review of the purely bosonic case}
\vskip .4truecm

\noindent 2.1 {\it Continuum approach}
\vskip .3truecm

\noindent Let us take the David-Distler-Kawai\ref{4} (DDK) description of $c<1$
conformal theories coupled to 2d gravity, where the partition function $Z$ on
the Euclidian sphere is defined by the following integral
over ``matter" ($X$) and Liouville ($\phi $) fields,
$$\displaylines{\hfill
Z=\int {{\cal D}_{\hat g}X\, {\cal D}_{\hat g}\phi \over V_{SL(2,C)}}\;
\E ^{-S}\quad ,\hfill (1)\cr
\hfill S={1\over 8\pi}\int {\rm d}^2w \shg  [\hat g^{ab} \partial
_a\phi \partial _b\phi+\hat g^{ab}\partial_aX\partial_bX -Q\hat R \phi
+2i\alpha_0\hat RX+8\pi\bar\mu \E^{\alpha \phi}]\; .
\hfill (2)\cr}
$$
$\hat g_{ab}$ as in (2) is the fiducial metric fixed by the conformal gauge,
$\hat R$ is the respective scalar curvature and $V_{SL(2,C)}$
is the volume of the residual $SL(2,\Complex)$
symmetry (a residual symmetry of the conformal gauge). The
parameter $\alpha _0$ and the background charge $Q$ are related to the matter
central charge $c$ by the relations
\vskip-7pt
$$
c=1-12\alpha _0^2\quad ,\qquad
Q=2\sqrt {2+\alpha _0^2}=\sqrt {{25-c\over 3}}\quad .
\eqno(3)
$$
The last term in Eq. (2), $\bar \mu \int {\rm d}^2w\; \sqrt {\hat g}\E ^{\alpha
\phi }$, measures the area of the spherical surface and is called  the
cosmological term. The coupling $\bar \mu $ is a bare cosmological constant
and the parameter $\alpha $ is chosen in such a way that $\E ^{\alpha \phi }$
has conformal dimension one,
$$
\Delta \left( \E^{\alpha\phi}\right) =-{1\over 2} \alpha
(\alpha +Q)=1\quad .
\eqno(4)
$$
The above equation has two solutions, namely $\alpha _\pm =-Q/2\pm |\alpha
_0|$. We choose $\alpha =\alpha _+\ge -Q/2$, which exhibits the correct
classical limit and also satisfies Seiberg's energy bound\ref{5}
($E(\alpha )=\alpha +Q/2\ge 0$).

As stated in the Introduction, we are interested in the scaling behaviour of
correlation functions of vertex operators. In the continuum approach we take
vertex operators ``dressed" by gravity, $V_i=\int {\rm d}^2z\; \sqrt {\hat
g}{\cal O}_i\E ^{\beta _i\phi }$, where ${\cal O}_i$ is the matter field or
vertex and the dressing factor $\E ^{\beta _i\phi }$ is determined by imposing
that ${\cal O}_i\E ^{\beta _i\phi }$ has conformal dimension $1$,
\vskip-8pt
$$
\Delta({\cal O}_i\E^{\beta_i\phi })=h_i-{1\over 2}\beta_i
(\beta_i + Q)=1\quad .
\eqno(5)
$$
Above, $h_i$ is the conformal dimension of ${\cal O}_i$. In particular, for a
tachyon ${\cal O}_i=\E ^{ik_iX}$, one finds $h_i={1\over 2}k_i(k_i-2\alpha _0)$
and therefore its dressing charge reads $\beta _i(k_i)=-Q/2+|k_i-\alpha _0|$.
Notice the degeneracy in this representation: the momenta $k_i$ and $2\alpha
_0-k_i$ imply the same conformal dimension and dressing. Therefore we may take
either
$$
T_k=\int {\rm d}^2z\sqrt {\hat g}\E ^{ikX+\beta (k)\phi }\qquad {\rm or}\qquad
T_{2\alpha _0-k}=\int {\rm d}^2z\sqrt {\hat g}\E ^{i(2\alpha _0-k)X+
\beta (k)\phi }
\eqno(6)
$$
to represent the same dressed operator, choosing one or the other according to
kinematic criteria. We call the above vertex ($T_k$) the tachyon, and $k$ is
its momentum. In particular the cosmological term or area operator
$T_0=\int {\rm d}^2z\; \sqrt {\hat g}\E ^{\alpha _+\phi }$ previously mentioned
may be as well represented by $T_{2\alpha _0}$$=\int {\rm d}^2z$ $\sqrt {\hat
g}$ $\E ^{2i\alpha _0X+\alpha _+\phi }$.

An $N$-tachyon correlation function ${\cal A}_N\equiv \langle T_{k_1}\cdots
T_{k_N}\rangle $ is  calculated, using Coulomb gas techniques. We choose the
fiducial metric $\hat g_{\alpha \beta }=\delta _{\alpha \beta }$. Integrating
over the matter field $X$ we obtain the momentum-conservation law
\vskip-8pt
$$
\sum _{i=1}^Nk_i=2\alpha _0 \quad ,
\eqno(7)
$$
which must be satisfied by every non-vanishing correlation function.

The gravitational contribution is more difficult to deal with, due to the
exponential self-interaction in the Liouville action.  At this  point we turn
to the zero-mode integration procedure\ref 6: defining
$\phi =\tilde \phi +\phi_0$,
where $\phi _0$ is the Liouville zero mode, we can explicitly integrate over
$\phi _0$ and use the free field propagator $\langle \E ^{\beta \tilde \phi
(z)}\E ^{\gamma \tilde \phi (w)}\rangle =(z-w)^{-\beta \gamma}$ for the
remaining field $\tilde \phi $.

After renormalizing the cosmological constant ($\bar \mu \to \mu$), we find the
$N$-point function
$$
{\cal A}_N = {\partial ^{N-3}\over \partial \mu^{N-3}} \mu ^{N+s-3}
\prod_{j=1}^N \Delta\left({1\over 2} (\beta_j^2-k_j^2)\right)
\sim \mu^s\quad ,\eqno(8)
$$
where
$\Delta (x)={\Gamma (1-x)\over \Gamma (x)}$, and
$s=-{1\over \alpha _+}\left( Q+\sum _{i=1}^N\beta _i\right) $.

After writing Eq. (8) we analytically continue the results, originally derived
for $s\in \Integer $, to $s\in \Complex $.
Moreover we constrain the kinematics choosing $N-1$ values of $k_i$ to be
bigger than $\alpha _0$ and one smaller (for a thorough discussion on the
subject see Refs. [7,10]). Notice that ${\cal A}_N$ factorizes in
terms of external leg contributions and its scaling behaviour with respect to
the cosmological constant $\mu $,
$
{\cal A}_N\sim \mu ^s=\mu ^{-Q/\alpha _+-\sum _{i=1}^N\beta _i
/\alpha _+}$,
provides the scaling dimension of the dressed vertex operators
$$
T_k=\int {\rm d}^2z\; \E ^{ikX+\beta (k)\phi }\quad \rightarrow \quad \mu
^{-\beta (k)/\alpha _+}\quad .
\eqno(9)
$$

The scaling of the partition function can also be taken from Eq. (8) and
defines
the so-called string susceptibility $\gamma $ as
$$
Z\sim {\cal A}_0\sim \mu ^{-Q/\alpha _+}=\mu ^{2-\gamma }\quad ,
\eqno(10)
$$
i.e. $\gamma =2+Q/\alpha _+$.

{}From the momentum conservation (7) we conclude that there is only one
tachyonic operator, with momentum $k=2\alpha _0$ and dressing charge $\beta
=\alpha _+$ (i.e. the area operator), whose expectation value is
non-vanishing,
$$
\langle T_{2\alpha _0}\rangle \sim \mu ^{-Q/\alpha _+ -1}\quad .
\eqno(11)
$$
The remaining 1-tachyon functions are expected to vanish.

The conservation law (7) also implies that  $T_{k_1}$ and $T_{2\alpha _0-k_1}$
represent the same operator, and we conclude that 2-point functions are
diagonal and scale as
$ \mu ^{-Q/\alpha _+-2\beta (k)/\alpha
_+} $.

In Ref.  [1] the authors used the wave functions of dressed vertex operators to
establish the correct Liouville/matrix model dictionary. The starting point was
the WdW equation\ref{1,5}:
$$
[H-\Delta_{\cal O}]\Psi _l({\cal O})=0 \quad ,
\eqno(12)
$$
where the wave function $\Psi _l({\cal O})$ corresponds to the insertion of a
dressed operator ${\cal O}$ on a surface of boundary $l$; $\Delta _{\cal
O}=1-h$ is the conformal weight of ${\cal O}$; $H$ is the Liouville Hamiltonian
$$
H={1\over 2}(\phi '+4\pi P)^2+Q(\phi '+4\pi P)'+{\mu \over 2\alpha _+^2}\E
^{\alpha _+\phi }+{Q^2\over 8} \quad .
\eqno(13a)
$$
In the minisuperspace approximation one ignores the space dependence and drops
derivative terms in $H$, which then reads
$$
H\equiv {1\over 2}p^2 + { \mu \over 2\alpha _+^2}\E^{\alpha _+ \phi} + { Q^2
\over 8}\quad ,\eqno(13b)
$$
with $p=-i\partial /\partial \phi $. Moreover, since the boundary length $l$ is
measured by the contour integral $l=\oint {\rm d}\xi \; \E ^{{1\over 2}\alpha
_+\phi
}$, we can associate
$$
{\partial \over \partial \phi }\quad \leftrightarrow \quad
{\alpha _+\over 2}\left( l{\partial
\over \partial l}\right)\qquad ,\qquad
{\mu \over 2\alpha _+^2}\E ^{\alpha _+\phi}\quad
\leftrightarrow \quad {\alpha _+^2\over 8}\mu l^2
\eqno(14)
$$
in which case Eq. (12) becomes the Bessel differential equation
$$
\left[ -\left( l{\partial \over \partial l}\right)^2 + \mu l^2 + \nu
^2\right]\Psi _l({\cal O})=0 \quad ,
\eqno(15)
$$
where
$
\nu ^2={Q^2-8\Delta \over \alpha _+^2}=\left( {Q\over \alpha _+}+2{\beta \over
\alpha _+}\right) ^2$.

Since the wave functions $\Psi _l({\cal O})$ are expected to decay in the
infrared (large $l$) limit, we take the modified Bessel functions of second
kind $\Psi _l({\cal O})\propto K_\nu (\sqrt \mu l)$ as appropriate solutions.
This result will be taken as a guide to unravel some tangles in the
continuum/discrete translation.

\vskip .5truecm
\noindent 2.2 {\it Discrete (matrix model) approach}
\vskip .3truecm

\noindent The Hermitian 1-matrix model is defined by the partition function
$$\displaylines{\hfill
Z=\int {\cal D}\phi \; \E ^{-{N\over \Lambda}\tr V(\phi )}\quad ,\quad \phi
^\dagger =\phi \quad ,\hfill (16)\cr
\hfill {\cal D}\phi = \prod _i{\cal D}\phi _{ii}\prod _{i<j}{\cal D}(\Re \phi
_{ij}){\cal D}(\Im \phi _{ij}) \quad ,\qquad
V(\phi )=\sum _k g_k\phi ^k \quad .\hfill\cr}
$$
One can generate different critical regimes by tuning the coupling constants
$g_k$. The potentials found by Kazakov in Ref.  [8] exemplify this property.
In the $m$-th critical regime the continuum limit of the model is
defined\ref{9} by a
double scaling limit, where the matrix size gets large ($N\to \infty $) and
the constant $\Lambda $ approaches a critical value ($\Lambda \to \Lambda _c$),
while the combination
$
N(\Lambda _c-\Lambda )^{1+1/2m}$
is kept constant. The resulting theory is described by the free energy
$$
{\cal F}=-{1\over \kappa ^2}\partial _{t_0}^{-2}u(t_n)\quad ,
\eqno(17)
$$
where $\kappa $ is the renormalized string coupling and $u$ is the specific
heat satisfying the {\it string equation}
$$
-t_0=\sum _{n=1}nt_nR_n[u]\quad ;\eqno(18)
$$
$R_n[u]$ are the Gel'fand-Dikii polynomials.
The dependence on the renormalized couplings $t_n$ is ruled by the KdV flows
$$
{\partial \over \partial t_n}u  = {\partial \over \partial t}R_{n+1}[u]
\quad .
\eqno(19)
$$
The exact $m$-th critical regime is defined by the limit
$$\eqalign{
&t_0\to t\quad ;\quad t_m\to -1\quad ;\cr
&t_n\to 0\quad ,\quad n\not =0, m\quad ,\cr}
\eqno(20)
$$
in which case the string equation becomes $mR_m[u]=t$.
Within the planar or spherical approximation, given by the limit $\kappa \to
0$, the Gel'fand-Dikii polynomials tend to $R_n[u]=u^n/n$, and the string
equation (18) becomes
$$
\sum _{n\ge 0} t_nu^n=0 \quad ,
\eqno(21)
$$
which in the $m$-th critical regime (20) implies $u^m=t$.

We associate a set of scaling operators $\sigma _n$ to the couplings $t_n$,
i.e. $\sigma _n \leftrightarrow {\partial \over \partial t_n}$. This frame of
operators and couplings was called the {\it KdV frame} by the authors in
Ref.  [1]. Correlation functions are described by the KdV flows, with the
following results in the planar approximation
$$
\left\langle \prod_{i=1}^n \sigma _{a_i}\right\rangle  = -{1\over
\kappa ^2}\partial _t^{n-2}{u^{a+1}\over a+1} \quad ,\quad a=\sum _{i=1}^n a_i
\quad .
\eqno(22)
$$
In the regime (20) we observe the  scaling behaviour of the free energy
as given by ${\cal F}\sim t^{2+1/m}=t^{2-\gamma }$,
which defines the susceptibility $\gamma _m=-1/m$. On the other hand the
$n$-point functions scale as $ t^{2+1/m+\sum (a_i/m -1)} $,
which therefore defines the scaling dimension $d_n=n/m$ of the operator $\sigma
_n$. Notice that neither any one of the 1- nor of the 2-point functions
vanish, as opposed to the
continuum results. Such differences obstruct a direct association
between vertex operators $T_k$ and the KdV frame of scaling operators $\sigma _
n$.

These difficulties can also be felt as one studies the wave functions of
scaling operators. We start defining macroscopic loop operators $W(l)$ of
finite length $l$, whose expectation value reads
$$
\langle W(l)\rangle ={1\over \kappa \sqrt {\pi l}}\int _{t_0}^\infty {\rm d}y\;
\E ^{-lu(y;t_n)}\quad .
\eqno(23)
$$
We suggest Ref.  [8] for a review on the subject. Notice that we can actually
integrate\ref{1} Eq. (23) by using the planar string equation (21),
finding as a result
$$
\langle W(l) \rangle ={1\over \kappa \sqrt \pi }\sum _{n\ge 0} t_nu^{n+1/2}\psi
_n(lu) \quad ,
\eqno(24)
$$
\vskip-10pt
\noindent with
$$
\psi _n(x)={1\over \sqrt x}\int _x^\infty {\rm d}z\, \left( {z\over x}\right)
^n\E ^{-z}=n!x^{-n-1/2}\E ^{-x}\sum _{s=0}^n{x^s\over s!} \quad .
\eqno(25)
$$
The functions $\psi _n(lu)$ display the dependence on the loop length $l$. They
are proportional to the wave functions associated to the insertion of scaling
operators, as follows
$$
\Psi _l(\sigma _n)=\langle \sigma _nW(l)\rangle ={\partial \over \partial
t_n}\langle W(l)\rangle \propto \psi _n(lu) \quad ,\eqno(26)
$$
i.e. Eq. (24) can be taken as an expansion of the loops in terms of wave
functions.
However we see that $\psi _n(lu)$ does not obey a Bessel equation, which
reinforces the differences between the KdV basis and the vertex operators.

In Ref.  [1] the authors resolved these discrepancies defining a new frame of
scaling operators $\widehat \sigma _n$ and the respective couplings
$\widehat t_n$,
which they called the {\it conformal field theory frame}.
They assumed that the differences between the observed correlation functions
should arise from contact terms. On the other hand a change of contact terms is
equivalent to an analytic redefinition of coupling constants. Thus one should
look for the proper set of couplings.

In order to change coupling constants and yet preserve scaling properties we
must examine their dimensions. From the way the loop length $l$ comes out in
\break
Eq. (23), we take the dimension of the specific heat, $[u]=[{\rm
length}]^{-1}$.
Thus the dimension of the couplings follow from the string equation (21),
$$
[t_n]=[{\rm length}]^{n-m}\quad .
\eqno(27)
$$\vskip-10pt
\noindent The ``physical" cosmological constant is usually taken to be the one
coupled to the area operator and should therefore have the dimension of inverse
of area or $[{\rm length}]^{-2}$. Since $[t]=[t_0]=[{\rm length}]^{-m}$ we
conclude that $t$ deserves the name of cosmological constant only when $m=2$.
In general $t_{m-2}$ is the coupling with the proper dimension of a
cosmological constant. If one wishes to compare results with the Liouville
approach one
should select another regime, different from the one defined by (20), where
scaling  properties are measured with respect to some coupling as $t_{m-2}$.

We can determine the conformal frame of couplings according to the following
criteria.
Consider an analytical transformation relating the KdV frame $\{ t_n\} $ to
another frame of couplings $\{ \widehat t_n\} $, with the general form
$$
t_n=C_n^i\widehat t_i +C_n^{ij}\widehat t_i\widehat t_j +\cdots
\eqno(28)
$$
In order to maintain scaling properties  (the operators $\widehat \sigma _n$
should be scaling fields as $\sigma _n$ already were), let us impose that the
analytic transformation  preserves the dimension of couplings, $[\widehat t_n]
= [t_n]$. In analogy to the limit (20) we shall define another $m$-th critical
regime in terms of the new set of couplings,
$$\eqalign{
&\widehat t_{m-2}\to \mu \quad ;\qquad \widehat t_m\to -1 \quad ;\cr
&\widehat t_n \to 0\quad ,\quad n\not =m-2,m \quad ,\cr }
\eqno(29)
$$
where $\mu $ is the ``physical" cosmological constant. The solution of the
string equation will become $u^2=\mu $ and correlation functions will scale as
functions of $\mu $. The operator $\widehat \sigma _{m-2}$ is expected to
correspond to the area or cosmological term.

Our immediate concern is to have 1- and 2-point functions compatible with the
Liouville predictions. In this case it is sufficient\ref{1} to consider lowest
(linear)-order terms in the transformation (28) around the critical point (29).
Thus it is convenient to shift the non-vanishing couplings as
$
\widehat t_{m-2}\to \mu +\widehat t_{m-2}$, $\widehat t_m \to -1 +
\widehat t_m $, so that the critical regime is achieved by the limit  $\widehat
t_n\to 0$ and we can deal with the $\widehat t_n$'s as perturbative couplings.

Considering these scaling restrictions and lowest-order approximations, the
transformation (28) is reduced to the form
$$
t_n=b_n\mu ^{(m-n)/2} + \sum _{s=0}^\infty a_s^{(n+2s)}\mu ^s\widehat t_{n+2s}
\quad ,
\eqno(30)
$$
where $b_n$ and $a_s^{(n)}$ are dimensionless coefficients to be determined.
However, the coefficients $b_n$ are not independent: they are given
by
$$
b_{m-2i}={1\over i}a_{i-1}^{(m-2)}\quad ;\qquad b_n=0\quad ,\quad n>m\quad .
\eqno(31)
$$
{}From Eq. (30) we derive ${\partial \over \partial \widehat t_n}=\sum _
{s=0}^{[n/2]}a_s^{(n)}{\partial \over \partial t_{n-2s}}$, which gives the
transformation of scaling operators
$$
\widehat \sigma_n = \sum _{s=0}^{[n/2]} a_s^{(n)} u^{2s} \sigma_{n-2s}
\quad .
\eqno(32)
$$
Therefore the coefficients $a_s^{(n)}$ uniquely characterize a given basis. In
order to find the conformal basis we can adopt two strategies, described below.

\vskip .3truecm
\noindent 2.2.a. {\it Minisuperspace approximation strategy}

\noindent In this case we use the wave functions as a guide\ref{1}
to determine the
coefficients $a_s^{(n)}$. From the Liouville description (see Eq. (15)) we
expect to find Bessel functions.
We take the expansion (24) for the 1-loop function, which holds in any regime,
and substitute the transformation (30). Recalling that $u^2\to \mu$ we find
$$\eqalign{
\langle W(l)\rangle &= {1\over \kappa \sqrt \pi }u^{m+1/2}\sum
_{s=0}^{[m/2]}b_{m-2s}\psi _{m-2s}(lu) \cr
&+{1\over \kappa \sqrt \pi }\sum _{n\ge 0}\widehat t_nu^{n+1/2}\left( \sum
_{s=0}^{[n/2]}a_s^{(n)} \psi _{n-2s}(lu)\right)\quad . \cr }
\eqno(33)
$$
Indeed we can obtain Bessel functions using
$$
K_{n+1/2}(x)= \sqrt{{\pi\over 2}} \sum_{s=0}^{[n/2]}
a_s^{(n)}\psi_{n-2s}(x)\qquad ,\quad
\eqno(34)
$$
with
$$
a_s^{(i)} = {(-1)^s\over 2^i} {(2i-2s)! \over s! (i-s)! (i-2s)!}\quad .
\eqno(35)$$
These are the coefficients we were looking for.
Using Eq. (31) we also find the corresponding expression for the remaining
coefficients
$$
b_{m-2s}={a_{s-1}^{(m-2)}-a_s^{(m)}\over (m-1/2)}\quad ,\quad
b_m=-{a_0^{(m)}\over (m-1/2)} \quad .
\eqno(36)
$$

Therefore the 1-loop function becomes an expansion in Bessel functions
$$\eqalign{
\langle W(l)\rangle &={\sqrt 2 \over \kappa \pi }{u^{m+1/2}\over (m-1/2)}\left[
K_{m-3/2}(lu) - K_{m+1/2}(lu) \right] \cr
&+{\sqrt 2\over \kappa \pi }\sum _{n\ge 0}\widehat t_n
u^{n+1/2}K_{n+1/2}(lu) \quad
,\cr }
\eqno(37)
$$
from which we obtain the wave functions
$$
\Psi _l(\widehat \sigma _n)=
\langle \widehat \sigma _nW(l)\rangle ={\sqrt 2\over \kappa \pi
}u^{n+1/2}K_{n+1/2}(lu) \quad ,
\eqno(38)
$$
which obey the Bessel equation (15) for $\nu =n+1/2$.
Notice also that in the exact regime (29) we have
$$\eqalign{
\langle W(l)\rangle &={\sqrt 2 \over \kappa \pi }{u^{m+1/2}\over (m-1/2)}\left[
K_{m-3/2}(lu) - K_{m+1/2}(lu) \right]\cr
&=-{2\over l}{\sqrt 2\over \kappa \pi
}u^{m-1/2}K_{m-1/2}(lu)=-{2\over l}\langle \widehat \sigma
_{m-1}W(l)\rangle \quad , \cr}
\eqno(39)
$$
i.e. $\langle \widehat \sigma _{m-1}W(l)\rangle = -{l\over 2} \langle
W(l)\rangle$, indicating\ref{1} that $\widehat \sigma _{m-1}$
is the boundary operator which measures the loop
length $l$.

We remark that the coefficients (35) of the conformal basis have an interesting
interpretation in terms of orthogonal polynomials: indeed from Eqs. (25) and
(34) we can write
$$
K_{n+1/2}(x)=\sqrt {\pi \over 2x}\int _x^\infty {\rm d}z\; P_n\!\!
\left( {z\over x}
\right) \E ^{-z} \quad ,
\eqno(40)
$$
where
$
P_n(x) =\sum _{s=0}^{[n/2]}a_s^{(n)}x^{n-2s}
$
are the Legendre polynomials. We shall understand the appearance of this family
of polynomials in the following subsection.

\vskip .3truecm
\penalty-1500
\noindent 2.2.b {\it Orthogonalization strategy}

\noindent The coefficients $a_s^{(n)}$ can also be determined by imposing that
the 2-point functions $\langle \widehat \sigma _i \widehat \sigma _j\rangle $
be
diagonal.
In fact, from the general transformation (32) and the 2-point functions of the
KdV frame $\langle \sigma _i\sigma _j\rangle $, which follow from Eq. (14), in
the regime $u^2\to \mu$, we find
$$
\langle \widehat \sigma_i \widehat \sigma_j\rangle
= -{\mu^{(i+j+1)/2}\over \kappa ^2}g_{ij}(a) \quad ,
\eqno(41)
$$
\vskip-10pt
\noindent where
$$
g_{ij} (a)=\sum _{s=0}^{[i/2]} \sum _{r=0}^{[j/2]}
{a_s^{(i)}a_r^{(i)}\over (i+j-2s-2r+1)}
\eqno(42)
$$
can be seen as a characteristic metric of the basis defined by some
coefficients $a_s^{(n)}$. Our aim is therefore to choose $a_s^{(n)}$ so that
$g_{ij}$ becomes diagonal. Observe that $g_{ij}$ satisfies
\vskip-9pt
$$
[1-(-1)^{i+j+1}] g_{ij}(a) = \int _{-1}^{+1} {\rm d} x P_i(x) \, P_j(x) \quad
,\quad P_i(x)= \sum _{s=0}^{[i/2]} a_s^{(i)} x^{i-2s} \quad ,
\eqno(43)
$$
where $P_i(x)=\sum _{s=0}^{[i/2]}a_s^{(i)}x^{i-2s}$ are characteristic
polynomials of a given basis. Thus our problem has been translated into
finding, among the characteristic polynomials, those which are orthogonal with
respect to the internal product $\langle P_i,P_j\rangle =\int _{-1}^{+1}{\rm
d}x\; P_i(x)P_j(x)$. The well-known solution is given by the Legendre
polynomials, whose coefficients are precisely those defined in (35).
Therefore the 2-point functions in the conformal frame read
$$
\langle \widehat \sigma _i\widehat \sigma _j\rangle =\cases{-{1\over \kappa ^2}
{\mu ^{i+1/2}\over 2i+1} \delta_{ij} &\quad ,\quad $i+j \; {\rm even}$\cr
{\scriptstyle {\rm {analytical\; in\; }}}\mu & \quad ,\quad $i+j\; {\rm odd}$
\cr }
\eqno(44)
$$
i.e. they are diagonal up to analytical terms in the coupling constants, as
expected from the Liouville approach.

Following any of the strategies we conclude that the conformal frame of
operators $\widehat \sigma _n$ is related to the KdV frame $\sigma _n$ by
$$
\widehat \sigma _n=\sum _{s=0}^{[n/2]}{(-1)^s\over 2^n}{(2n-2s)!\over
s!(n-s)!(n-2s)!}\mu ^s \sigma _{n-2s} \quad ,
\eqno(45a)
$$
which can also be inverted
$$
\sigma _n=\sum _{s=0}^{[n/2]}{\sqrt \pi n!\over 2^n}{(2n-4s+1)\over s!\Gamma
(n-s+3/2)}\mu ^s\widehat \sigma _{n-2s} \quad .\eqno(45b)
$$

We must also test the 1-point functions. From Eqs. (44) and (45$b$) we can
calculate ${\partial \over \partial \mu }\langle \sigma _n\rangle =\langle
\widehat
\sigma _{m-2}\sigma _n\rangle $ which, after integration, gives
$$
\langle \sigma _{m-2+2s}\rangle =-{\sqrt \pi \over \kappa ^2}{\mu
^{m+k-1/2}\over 2^{m-2+2k}}{(m-2+2k)!\over k!\Gamma (m+k+1/2)}\quad .
\eqno(46)
$$
Now we can take Eq. (45$a$) and easily calculate the 1-point functions $\langle
\widehat \sigma _n\rangle $. It turns out that the only non-vanishing cases are
$$
\langle \widehat \sigma _{m-2} \rangle  = -{4\over \kappa ^2}
{\mu^{m-1/2}\over (2m-1)(2m-3)}\quad ,\quad
\langle \widehat \sigma _{m} \rangle  = +{4\over \kappa ^2}
{\mu^{m+1/2}\over (2m+1)(2m-1)}\quad .\eqno(47a,b)
$$
This is also in agreement with the Liouville approach: $\langle \widehat \sigma
_{m-2}\rangle $ is the average area, as in Eq. (11); on the other hand
$\langle \widehat \sigma _m\rangle $ can be interpreted\ref{1} as a
non-tachyonic part of the  average energy.
Since $\langle \widehat \sigma _{m-2} \rangle ={\partial \over \partial \mu }
{\cal F}$, we can integrate Eq. (47$a$) and obtain the scaling behaviour of the
free energy in this regime: ${\cal F}\sim \mu ^{m+1/2}$.
Higher-order correlation functions would require the analysis of higher-order
terms in the transformation (30). Nevertheless we know that, within the regime
(29), the $N$-point functions must scale as
$$
\left\langle \prod_{i=1}^N \widehat \sigma _{n_i} \right\rangle   \sim
\mu^{{m+1/2}+ \sum_i {(n_i-m)\over 2}}\quad ,
\eqno(48)
$$
which is the proper expression to be compared with Eq. (8).

As concerns the string equation one comment is in order. In the regime
(29) the string equation (21) becomes
$$
\sum _n b_n\mu ^{(m-n)/2}u^n={\mu ^{m/2}\over (m-1/2)}\left[ P_{m-2}\left(
{u\over \sqrt \mu }\right) - P_m\left( {u\over \sqrt \mu }\right) \right] =0
\quad .
\eqno(49)
$$
Recalling that Legendre polynomials satisfy $P_n(1)=1$ we conclude that the
solution $u=\sqrt \mu$ has been consistently chosen.

\vskip .3truecm
2.3 {\it Continuum/discrete dictionary}
\vskip .3truecm

\noindent
Suppose that the $m$-th critical Hermitian 1-matrix model corresponds to a
$(p,q)$-minimal model coupled to gravity. In order to determine $p$ and $q$ we
compare scaling exponents in the $t$ and $\mu $ regimes.

In the continuum approach the partition function scales as $Z\sim \mu
^{-Q/\alpha _+}$ as stated in Eq. (10). On the other hand we learned
that $t$ should be the coupling of the minimal-weight operator, so that in a
$t$-type regime we expect the partition function to scale as $Z\sim t^{-Q/\beta
_{min}}$, where $\beta _{min}$ is the dressing charge of the minimum weight
operator. These exponents can be written in terms of $p$ and $q$,
$$
-{Q\over \beta _{min}}={2(p+q)\over (p+q)-1} \qquad, \qquad
-{Q\over \alpha _+}=1+{q\over p} \quad .
\eqno(50)
$$

Both cases were also studied in the discrete $m$-th critical model, where
the free energy scales as ${\cal F}\sim t^{2+1/m}$ and ${\cal F}\sim \mu
^{m+1/2}$, in the $t$ and $\mu $ regime respectively.
Identifying the scaling exponents from each approach we find two equations,
$$\eqalign{
{2(p+q)\over (p+q)-1}&=2+{1\over m}\quad ,\cr
1+{q\over p}&=m+{1\over 2}\quad ,\cr }
\eqno(51)
$$
which uniquely determine $(p,q)=(2,2m-1)$.
\vskip .2truecm

Next we examine the scaling operators. In the $(p,q)$ model there are ${1\over
2}(p-1)(q-1)$ primary operators, usually labelled ${\cal O}_{rr'}$, $1\le
r\le q-1$ and $1\le r' \le p-1$, with conformal weights $\Delta
_{rr'}={(rp-r'q)^2-(q-p)^2\over 4pq}$. With this notation the index $\nu $ that
characterizes the wave function $\Psi _l({\cal O}_{rr'})$  is given by
$$
\nu =\sqrt {Q^2-8\Delta _{rr'}\over \alpha _+^2}=r'{q\over p}-r\quad .
\eqno(52)
$$
For $(p,q)=(2,2m-1)$ we have $r'=1$ and $1\le r\le m-1$ (we can ignore the
range
$m\le r\le 2(m-1)$, which only doubles the spectrum of weights). Therefore $\nu
=m-r-1/2$. On the other hand the discrete model predicts $\nu =n+1/2$. We
conclude that $r=m-1-n$, with
$n=0,\cdots ,m-2$, spans the set of primary operators
and we can finally draw the identification\ref{1}
$$
\widehat \sigma _n\quad \leftrightarrow \quad \int {\rm d}^2\xi \; \E^{\beta
_n\phi }{\cal O}_{m-1-n,1}\quad ,\quad n=0,1,\cdots ,m-2\quad .
\eqno(53)
$$
Concerning the remaining scaling operators, we have two special cases:
as indicated by Eq. (39), $\widehat \sigma _{m-1}$ corresponds to the boundary
operator
$$
\widehat \sigma _{m-1}\leftrightarrow \oint {\rm d}\xi \;\E^{\alpha
_+\phi /2}\quad ,
\eqno(54)
$$
while $\widehat \sigma _m$ is part of the energy operator, as suggested in
Ref.  [1]:
$$
\widehat \sigma _m-{(2m-3)\over (2m+1)}\mu \widehat \sigma _{m-2}\;
\leftrightarrow \; {\rm energy}\quad .
\eqno(55)
$$
For the operators
$\widehat \sigma _n$, $n>m$, we verify that
their conformal weights differ from the weights of $\widehat
\sigma _n$, $n\le m$, by integers, and thus correspond to secondary
operators.
The 1- and 2-point functions found in the discrete and continuum approaches are
in perfect agreement. Higher-order correlators involve the issue of fusion
rules and lay beyond the approximation assumed in Eq. (10). Nevertheless the
scaling factors agree with the predictions of the continuum.

Now we are ready to generalize the previous computations for the non-critical
superstring.
\vskip .5truecm
\penalty-1500
\noindent 3. {\bf Supersymmetric non-critical strings}
\vskip .3truecm

\noindent 3.1 {\it Super-Liouville approach to non-critical superstrings}
\vskip.3truecm

\noindent The results in this section have been obtained in Ref.  [2], using
the formulation of Ref.  [11] (see also the review [10]).
The total supersymmetric action is given by
$S=S_{SL}+S_M+S_{gh}$, where
$$
S_{SL}=\int {{\rm d}^2z{\rm d}^2\theta \over 4\pi }\; E\left( {1\over
2}D_\alpha \Phi _{SL}D^\alpha \Phi _{SL}-{Q\over 2}Y\Phi _{SL}-4i\overline
\mu\E ^{\alpha _+\Phi _{SL}}\right)
\eqno(56)
$$
is the super-Liouville action, representing the supergravity sector of the
superstring in the superconformal gauge; $\Phi _{SL}= \phi +\theta \psi +\bar
\theta \bar \psi +\theta \bar \theta F$ is the Liouville superfield; $\overline
\mu $ is a bare cosmological constant; $E$ is the superdeterminant of the
super-zweibein and $Y$ the curvature superfield\ref{2,10,11}. Also,
$$
S_M=\int {{\rm d}^2z{\rm d}^2\theta \over 4\pi }\; E\left( {1\over 2}D_\alpha
\Phi _MD^\alpha \Phi _M-i\alpha _0Y\Phi _M\right)
\eqno(57)
$$
corresponds to the $c<3/2$ supermatter action in the Coulomb gas formulation,
with matter superfield $\Phi _M=X+\theta \zeta +\bar \theta \bar \zeta +\theta
\bar \theta G$. The term $S_{gh}$ stands for the ghost action whose explicit
form we do not need here.
The background charge $\alpha _0$ is related to the matter central charge:
$c={3\over 2}(1-8\alpha _0^2)$.
To compute $Q$ and $\alpha_+$ we have to impose respectively that the
total central charge $c_T$ vanishes and $\E^{\alpha_+\Phi_{SL}}$ be a
$(1/2,1/2)$ conformal operator.
We need the bosonic piece of the super energy momentum tensor
which is given in components by
$$
\eqalignno{
T_{SL }&= - {1\over 2}\colon \partial \phi\partial \phi\colon - {1\over 2}
\colon \psi\partial \psi\colon + {Q\over 2}\partial ^2 \phi \quad ,\quad
T_{M }= - {1\over 2}\colon \partial X\partial X\colon - {1\over 2}\colon\zeta
\partial \zeta \colon - i\alpha_0 \partial ^2X  &\cr
T_{gh} &= T_{bc} + T_{\beta \gamma} = \colon c \partial b \colon + \colon
2(\partial c)b\colon - {3\over 2}\colon (\partial \gamma)\beta \colon -{1\over
2}\colon\gamma\partial\beta\quad , &(58)\cr}
$$
where $b,c$ $(\beta,\gamma)$ are ghost fields.
We use free propagators for all fields. The central charges for super-Liouville
and ghosts are computed in the usual way, that is,
$c_{SL} = {3\over 2}(1+2Q^2)$ and $c_{gh} =-15$.
Imposing that  $c_{SL} + c + c_{gh}=0$ we deduce that
$ Q = 2\sqrt{\alpha^2_0+1}$.
The conformal weight $\Delta$ of an operator $\E^{\alpha\Phi}$ is
$ \Delta(\E^{\alpha\Phi}) = - {\alpha(\alpha+Q)\over 2}.$
Requiring that $\Delta = 1/2$ we get
$\alpha_\pm = -{Q\over 2} \pm {1\over 2}\sqrt{Q^2-4} = - {Q\over
2} \pm \vert \alpha_0\vert$, with $ \alpha_+\alpha_-=1$.

The particle content consists of a scalar (Neveu-Schwarz, NS) and a spinor
(Ramond, R) vertex, both massless.
The Neveu-Schwarz vertex is the supersymmetric extension of the tachyon,
$$\eqalign{
\psi_{{}_{NS}}(k)\!
&=\! \int\! {\rm d}^2z{\rm d}^2\theta \; \E ^{ik\Phi _M+\beta \Phi _{SL}}\cr
&=\! \int\! {\rm d}^2 z \E^{ikx + \beta \phi}[(ik\zeta + \beta
\psi)(ik\overline
\zeta + \beta \overline \psi) + \beta F + ikG]\quad .\cr }
\eqno(59)$$
As before, we impose that $\E^{ik\Phi_M +
\beta \Phi_{SL}}$ be a $(1/2,1/2)$ conformal operator,
$$
\Delta(\E^{ik\Phi_M + \beta\Phi_{SL}}) = {1\over 2}k(k-2\alpha_0) - {\beta\over
2}(\beta + Q) = {1\over 2}\quad ,
\eqno(60)
$$
which fixes the NS dressing charge;  as  in the bosonic case, $
\beta (k)=-{Q\over 2}+\vert k-\alpha _0\vert$.

The auxiliary fields $F$ and $G$ appear in a trivial way in the action.
Their propagators consist of delta functions,
$
\langle F(z_i)F(z_j) \rangle =\langle G(z_i)G(z_j)\rangle\sim \delta^{(2)}
(z_i-z_j)
$.
The contractions of other fields typically give
$
\langle{\E^{a\phi(z_i)}\E^{b\phi(z_j)}}\rangle\!\sim \!\vert
z_i\!-\!z_j\vert^{^{-2ab}}
$.
This problem will be circumvented by discarding such fields,
i.e. we fix $F=G=0$. This amounts to assuming $ab<0$\ref{12}. After such
simplification the NS vertex becomes
$$
\psi_{{}_{NS}}(k) = \int \!{\rm d}^2 z {\rm d}^2 \theta \;\E^{ik\Phi_M + \beta
\Phi_{SL}}
= \int {\rm d} z (ik\zeta + \beta \psi) \int {\rm d} \overline z (ik\overline
\zeta + \beta \overline \psi) \E^{ikx+\beta\phi}\; .
\eqno(61)
$$

In order to calculate correlation functions we must consider the residual
$OSP$ $(2,1)$ symmetry of the superconformal gauge. After choosing
$
\widetilde z_1=0$, $\widetilde z_2=1$, $\widetilde z_3=
\infty$, $\widetilde \theta _2=\widetilde \theta_3=0
$ and renormalizing the cosmological constant $\overline \mu \to \mu $, one
finds\ref{2}
$$
{\cal A}_N= {\partial ^{N-3}\over \partial \mu^{N-3}}\mu^{N+s-3} \prod_{j=1}^N
\Delta\left({1\over 2} + {1\over 2}(\beta_j^2-k_j^2)\right)\quad ,
\eqno(62)
$$
with $s=-(Q+\sum _i\beta _i)/\alpha _+$, i.e. the results are completely
similar to the bosonic case (recall Eq. (8)).

For the Ramond vertex we have to consider other fermionic contributions.
First we bosonize the fermions $\psi $ and $\zeta $ into a bosonic massless
field $h$ in the usual way, that is\ref 7
$$
\psi \pm i\zeta =\sqrt 2 \E^{\pm ih}
\eqno(63)
$$
with the contraction $\langle h(z)h(w)\rangle =-{\rm ln}(z-w)$ and the
superselection rule
$$
\left\langle \sum _i\E ^{q_ih(z_i)}\right\rangle \not =0 \quad {\rm iff} \quad
\sum _iq_i=0\quad .
\eqno(64)
$$
Because of supersymmetry the Ramond vertex must also contain a ghost spinor
field $\Sigma $ whose bosonized form $\Sigma _{\pm 1/2}=\E ^{\pm \sigma /2}$
involves a massless bosonic field $\sigma $ with propagator $\langle \sigma
(z)\sigma (w)\rangle =-{\rm ln}(z-w)$. The two solutions have different
conformal dimensions, $\Delta (\E ^{-\sigma /2})=3/8$ and $\Delta (\E ^{\sigma
/2})=-5/8$. The requirement
$$
\Delta\left(\E^{{i\over 2}\epsilon h(z) +ikX(z)+\beta \phi (z)}\Sigma
(z)\right) ={1\over 8} +{1\over 2}k (k-2\alpha_0)-{1\over 2}\beta (\beta +Q)
+\Delta \left(\Sigma\right) =1
\eqno(65)
$$
is enough to determine the proper choice. Indeed
the Ramond vertex $V_{_R}(k,\epsilon )$ should represent a massless particle.
The on-shell condition $(k-\alpha_0)^2-(\beta+Q/2)^2=0$, equivalent to
$E^2-p^2=0$, implies that $\Delta \left(\Sigma\right)=3/8$, which selects
the  solution $\Sigma =\Sigma_{-1/2}=\E^{-\sigma/2}$.
Therefore our spinor vertex can be written as
$$
V_{_R}(k,\epsilon )=V_{-{1\over 2}}(k,\epsilon )=\int\! {\rm d}^2z\,
\E^{-{1\over 2}\sigma (z)+{i
\over 2}\epsilon h(z) +ikX(z)+\beta \phi (z)} \quad .
\eqno(66)
$$
{}From the Dirac equation we find for the dressing the expression\ref{7} $\beta
(k,\epsilon )=-{Q\over 2} +\epsilon (k-\alpha_0)$.
There are further versions of these operators obtained by a certain
procedure defined in Ref.  [13], leading to the so-called pictures of
vertices.
We remark that the field $\sigma $ has background charge $-2$, which implies
another superselection rule,
$$
\left\langle \sum _i\E ^{q_i\sigma (z_i)}\right \rangle \not =0 \quad {\rm iff}
\quad \sum _i q_i=-2\quad .
\eqno(67)
$$

Another useful picture of the NS vertex is the following
$$
\psi_{_{NS}}^{^{(-1)}}(k)=\int \!{\rm d}^2z \, \E^{-\sigma(z)+ikX(z)+\beta
\phi(z)}\quad .
\eqno(68)
$$
The above vertex is BRST-invariant and can be {\sl ``picture-changed"
 } into $\psi_{NS}=$ $\left[ Q_{BRST},\xi\psi_{NS}^{(-1)}\right]
$, where $\xi$ is the ghost zero mode (see [13] for details). This is however
not BRST-exact.

A mixed $N$-point correlator
$$
{\cal A}_N^{(n,N-n)}=\left\langle \prod _{i=1}^n V_{-{1\over 2}}(k_i,\epsilon
_i)\prod _{j=n+1}^N\psi _{_{NS}}(k_j)\right\rangle
\eqno(69)
$$
is obtained by integrating over matter ($X_0$) and Liouville ($\phi _0$)
bosonic
zero modes. The total momentum conservation law (7) still holds, supplemented
by the rules (64) and (67). As in the bosonic case we find a factorizable
result\ref{2}
$$
{\cal A}_N^{(n,N-n)} \sim \mu^s \prod_{i=1}^n \Delta\left({1\over 2}(\beta_i^2
- k_i^2)\right) \prod_{j=n+1}^N \Delta \left( {1\over 2} + {1\over 2}(\beta_j^2
- k_j^2)\right) \quad ,
\eqno(70)
$$
where the exponent $s$, defined in (8), now includes spinor vertex momenta.
Notice that the inclusion of the vertices $V_{-{1\over 2}}$ does not alter the
scaling of the partition function and Eq. (10) still holds. The insertion of a
NS or R vertex modifies the scaling behaviour by a factor $\mu ^{-\beta/\alpha
_+}$, $\beta $ being the corresponding dressing charge.

The bosonic case has taught us how important the analysis of 1- and 2-point
functions are in the comparison between continuum and discrete results. So they
must be in the supersymmetric theory.
As in Eq. (11), momentum conservation and the extra rules (64) and (67) imply
that the ``area" operator $\psi _{_{NS}}(2\alpha _0)$ is the only vertex with
non-vanishing expectation value, scaling as
$$
\langle \psi _{_{NS}}(2\alpha _0)\rangle \sim \mu ^{-Q/\alpha _+-1}\quad .
\eqno(71)
$$
In the NS sector we have an orthogonality property similar to the one valid
before for the operators (6), namely
$$
\langle \psi _{_{NS}}(k)\psi _{_{NS}}(2\alpha _0-k)\rangle \sim \mu ^{-Q/\alpha
_+-2\beta (k)/\alpha _+}\quad .
\eqno(72)
$$

Concerning the R sector, the superselection rule (64) implies that every
1-point function $\langle V_{-{1\over 2}}\rangle $ must vanish.
{}From Eq. (70) and the rules (64) and (67) we also conclude that
${\cal A}_2^{(2,0)}=0$. However, considering
the picture-changed NS vertex (68) we can build a non-vanishing 2-spinor
function, $\langle V_{-{1\over 2}}(k,\epsilon )V_{-{1\over 2}}(2\alpha
_0-k,-\epsilon )\psi _{_{NS}}^{^{(-1)}}(0)\rangle \not =0$. The operator
$\psi _{_{NS}}^{^{(-1)}}(0)=\int\! {\rm d}^2z\, \E ^{-\sigma +\alpha _+\phi }$
acts as a screening required by the rule (67), its ``engineering" dimension
is $[{\rm length}]^2$ and its coupling constant must tend to $\mu ^2$ in the
$\mu $-regime. The screening vertex $\mu ^2\psi
_{_{NS}}^{^{(-1)}}(0)$ thus defined contributes with a scaling factor $\mu
^{-\alpha _+/\alpha _++2}=\mu $. We therefore have the following screened
2-point function
$$
\langle V_{-{1\over 2}}(k,\epsilon )V_{-{1\over 2}}(2\alpha
_0-k,-\epsilon )\mu ^2\psi _{_{NS}}^{^{(-1)}}(0)\rangle \sim \mu ^{-Q/\alpha
_+-2\beta /\alpha _++1}
\quad .
\eqno(73)
$$
We could interpret the above result in the following way: the spinor vertex
$V_{-{1\over 2}}(k,\epsilon )$ contributes with the scaling factor $\mu
^{-\beta /\alpha _+}$, while its ``conjugate" vertex in this representation is
the composite operator $\mu ^2 \psi _{_{NS}}^{^{(-1)}}(0)\times V_{-{1\over
2}}(2\alpha _0-k,-\epsilon )$, which scales as $\mu ^{-\beta /\alpha
_++1}$. This interpretation will be further supported by the discrete model
results.

A delicate issue concerning the comparison with the matrix models results is
the definition of the matter sector of the Ramond vertex, since the
gravitational and matter fermions merge into $\E^{{i\over 2}h}$, and an
identification of matter contribution with the Kac table results turns out
to be difficult.
Recall that in general for a Ramond vertex
$
{3\over 8}+ {1\over 8} + {1\over 2}k(k-\alpha_0) - {1\over 2}\beta (\beta+Q)=1
$,
where the $3/8$ comes from the ghost contribution,
and $1/8$ from  $\E^{{i\over 2}h}$.
Therefore, since the latter is neither pure matter nor pure
gravity, we have to disentangle their contributions and find the relation
between the pure matter conformal dimension $\Delta_{ Kac} \, (\Delta_K)$
and the dressing.
Gravitational dressing gives a contribution $-{1\over 2}\beta
(\beta +Q)+ {1\over 16}$, as argued in Ref.  [11].
(The Neveu-Schwarz field is expanded in half-integer components
and displays no zero mode, while  the Ramond field is expanded in integer
modes: the zero mode contributes $1/16$). Put in another way, in order to
maintain the form of the Virasoro algebra we must have, in the Ramond sector,
 shift $L_0$ into $L_0 + {1\over 16}$, where $1/16$ is the ground-state energy.
Therefore we have,
$
\Delta _K = \Delta + {1\over 16}
$,
where $\Delta = {1\over 2}k(k-\alpha_0)$.

\vskip .5truecm
\noindent 3.2 {\it Discrete super-eigenvalue model}
\vskip .3truecm

\noindent In Ref.  [3] a discrete model was proposed as a supersymmetric
extension of the effective eigenvalue theory, which in its turn comes from the
angular integration of the Hermitian 1-matrix model. It is  defined by the
supersymmetric partition function
$$\eqalign{
&Z_s=\int \prod _{n=1}^N{\rm d}\lambda _n{\rm d}\theta _n\, \E ^{-{N\over
\Lambda }\sum _i V_s(\lambda _i,\theta _i)} \prod _{i<j}^N (\lambda _i-\lambda
_j-\theta _i\theta _j) \quad ,\cr
&V_s(\lambda _i,\theta _i)=\sum _{k=0}^\infty (g_k\lambda _i^k+\xi
_{k+1/2}\theta _i\lambda _i^k) \quad ,\cr }
\eqno(74)
$$
where $\lambda _i, \theta _i$ are commuting and anticommuting eigenvalue
variables, respectively, and $g_k, \xi _{k+1/2}$ are the commuting and
anticommuting coupling constants.

The planar superloop equations and the double scaling limit of this model were
studied in Refs. [3,14]. Higher genera were also considered in Ref.  [14] but
it
was only after Ref.  [15] that a non-perturbative definition was given in terms
of a supersymmetric KdV hierarchy.

To compare results with the continuum approach the planar approximation will
suffice. The $m$-th critical double scaling limit is similar to the bosonic
case. For even bosonic potentials the resulting theory is
described\ref{3} by the supersymmetric free energy
$$
{\cal F}_s=-{1\over 2\kappa ^2}\partial _t^{-2}[1-\partial _t\tau _+\tau
_-\partial _t]u \quad ,
\eqno(75)
$$
where the functions $u$ and $\tau_\pm $, respectively bosonic and fermionic,
satisfy the equations
$$
t=u^m-\sum _{n\ge 0}t_nu^n\qquad ,\qquad
\tau _\pm(u) =\sum_{n\ge 0}\tau _n^\pm u^n \quad \quad ,
\eqno(76)
$$
in terms of the renormalized bosonic (fermionic) coupling constants
$t_n$ ($\tau _n^\pm $). It is also convenient to define the functions
$
\rho_n[u] \equiv(1-\partial _t\tau _+\tau _-\partial _t){u^{n}\over n}
$, generalizing the monomials $R_n[u]=u^n/n$ of the bosonic theory, whose
flows are given by
$$
{\partial \over \partial t_n}\rho _m[u] ={\partial \over
\partial t}\rho _{n+m}[u]\qquad ,\qquad
{\partial \over \partial \tau _n^\pm }\rho _m[u] ={\partial
\over \partial \tau _\pm }\rho _{n+m}[u]\quad .
\eqno(77)$$
Using the string equation (23) we can also write the functions $\rho _n$ as
$$
\rho _n[u]=\left( 1-D_+D_-\right) {u^n\over n}\quad ,
\eqno(78a)
$$
\vskip-10pt
\noindent where
$$
D_\pm \equiv \sum _{i\ge 0}\tau _i^\pm {\partial \over \partial t_i}\quad .
\eqno(78b)
$$
In this case we can make explicit the quadratic dependence\ref{3,15} of
fermionic  couplings exhibited by the free energy:
$$
{\cal F}_s=(1-D_+D_-)\left[ -{1\over 2\kappa ^2}\partial _t^{-2}u\right]
={1\over 2}(1-D_+D_-){\cal F}_{bosonic}
\quad .\eqno(79)
$$

We associate scaling operators to the couplings,
$\sigma_n\leftrightarrow {\partial\over\partial t_n} $,
$\nu^\pm_n\leftrightarrow {\partial\over\partial t_n}$
and their correlation functions follow easily
$$
\eqalignno{
\left\langle \prod _{i=1}^n\sigma _{a_i}\right\rangle &=-{1\over 2\kappa ^2}
\partial_t^{n-2}\rho _{a+1}[u]\quad ,\quad a=\sum _i a_i\quad ,&(80a)\cr
\left\langle \nu _k^\pm
\prod _{i=1}^n\sigma _{a_i}\right\rangle &=-{1\over 2\kappa ^2}\partial
_t^{n-2}{\partial \over \partial \tau _\pm }
\rho _{k+a+1}[u]\quad ,&(80b)\cr
\left\langle \nu _k^+\nu _l^-
\prod _{i=1}^n\sigma _{a_i}\right\rangle &=-{1\over 2\kappa ^2}\partial
_t^{n-2}{\partial \over \partial \tau _+}{\partial \over \partial \tau _-}
\rho _{k+l+a+1}[u]\quad .&(80c)\cr }
$$
In analogy to the bosonic case and considering the results of Ref. [15] we
shall
call this set of operators/couplings the {\it super-KdV frame}.
Now we can compute correlation functions in the $t$-regime given by the limits
$t_n,\tau _n^\pm \to 0$. The free energy scales as in
the bosonic model,
$$
{\cal F}\sim t^{2+1/m}\Rightarrow \gamma=-1/m
\eqno(81)
$$
while higher-order correlation functions scale as\ref{3}
$$
\eqalignno{
\left\langle \prod _{i=1}^n\sigma _{a_i}\right\rangle &\sim t^{2+{1\over
m}+\sum
({a_i\over m}-1)}\quad , &(82)\cr
\left\langle \nu _k^+\nu _l^-\prod_{i=1}^n \sigma_{a_i}\right\rangle&\sim
t^{2+{1\over m} +({k\over m}-{1\over 2m}-1)+({l\over m}+{1\over 2m}-1)+\sum
({a_i\over m}-1)},&(83)\cr}
$$
implying the scaling dimensions $d_{\sigma _n}={n\over m}$ and
$d_{\nu^\pm_n}={n\over m}\mp {1\over 2m}$. Further correlators vanish.

We can foresee difficulties with the super-KdV basis similar to the bosonic
case: indeed we have infinitely many non-vanishing 1-point functions,
$\langle\sigma_i\rangle \not =0$, while the 2-point functions
$\langle\sigma_i\sigma_j\rangle$  and
$\langle\nu_i^+\nu_j^-\rangle$ are not diagonal. The results (71) to (73) show
that we cannot associate the operators ($\sigma _i,\nu
_i$) to ($\psi _{_{NS}},V_{-{1\over 2}}$) in a simple way.

\vskip.5truecm
\penalty-1500
\noindent 3.3 {\it Superconformal frame of scaling operators}
\vskip .3truecm

\noindent Let us begin to investigate the dimension of the coupling constants.
As in the bosonic case the specific heat $u$ has dimension of inverse
length, $[u]=[{\rm length}]^{-1}$. From the string equation and the
renormalization procedure used in Ref.  [3] we verify that
$$
\left[ t_n\right]  =\left[ {\rm length}\right]^{n-m}\quad ,\qquad
\left[ \tau_n^\pm\right] =\left[ {\rm length}\right]^{n-m\mp{1\over 2}}\quad .
\eqno(84)
$$
Recalling the super-Liouville action (56), we observe that the ``area" operator
actually has the dimension of a length. In this case the ``physical"
cosmological constant $\mu $ has dimension of inverse length. We conclude
that in general $t_{m-1}$ is the coupling with the proper dimension of a
cosmological constant rather than $t$. The correct comparison with the
continuum results requires that we replace the $t$-regime by a $\mu $-like
scaling regime.

As outlined in the bosonic case, we take the string equation (21), consider an
analytical transformation from the super-KdV basis $\{ t_n, \tau _n^\pm \} $ to
a generic one $\{ \widehat t_n, \widehat \tau _n ^\pm \} $, and replace the
$t$-regime (20) by the following one:
$$\eqalign{
&\widehat t_{m-1}\to \mu \quad ;\qquad \widehat t_m\to -1 \quad ;\cr
&\widehat t_n \to 0\quad ,\quad n\not =m-1,m \quad ;\qquad
\widehat \tau _n^\pm \to 0 \quad ,\quad \forall n\quad ,\cr }
\eqno(85)
$$
with solution $u=\mu $ for every $m$-th critical model.
Since we are going to concentrate on the properties of 1- and 2-point functions
we must make an extra shift of the non-vanishing couplings, $\widehat
t_{m-1}\to \mu +\widehat t_{m-1}$ and $\widehat t_m\to -1+\widehat t_m$, and
then treat $\widehat t_n, \widehat \tau _n^\pm $ as perturbative couplings.
Imposing also  that the dimensions are preserved,  i.e. $[\widehat t_n]=[t_n]$
and $[\widehat \tau _n^\pm ]=[\tau _n^\pm ]$, the lowest-order transformations
must have the form
$$
t_n=B_n\mu ^{m-n}+\sum _{s=0}^\infty A_s^{(n+s)}\mu ^s \widehat t_{n+s} \quad ,
\eqno(86)
$$
\vskip-9pt
\noindent with
$$
B_{m-i}={1\over i}A_{i-1}^{(m-1)}\quad ;\qquad B_n=0 \quad ,\quad n>m
\eqno(87)
$$
\vskip-9pt
\noindent and
$$
\tau _n^\pm =\sum _{s=0}^\infty A_s^{(n+s)}\mu ^s \widehat \tau _{n+s}^\pm
\quad .
\eqno(88)
$$
The critical regime is therefore defined by the limit $\widehat t_n, \widehat
\tau _n^\pm \to 0$.
As we shall see, the choice of the same coefficients $A_s^{(n)}$ in both
transformations (86) and (88) implies that $\langle \widehat \sigma _i \widehat
\sigma _j\rangle $ and $\langle \widehat \nu _i^+\widehat \nu _j^-\rangle $ are
simultaneously diagonalized and also preserves supersymmetry.

{}From Eqs. (86) and (88) we obtain the new set of scaling operators,
$$
\widehat \sigma _n= \sum _{s=0}^n A_s^{(n)}\mu ^s\sigma _{n-s}\quad ,\qquad
\widehat \tau _n^\pm =\sum _{s=0}^n A_s^{(n)}\mu ^s\tau _{n-s}^\pm \quad .
\eqno(89)
$$
In order to find the proper coefficients $A_s^{(n)}$ we shall follow the
orthogonalization procedure described in the bosonic model. From the
transformation (89) and the general correlators given in (80) we find
$$
\langle \widehat \nu _i^+\widehat \nu _j^-\rangle =
\langle \widehat \sigma _i\widehat \sigma _j\rangle
=-{1\over 2\kappa ^2}\mu ^{i+j+1}g_{ij}(A) \quad ,
\eqno(90)
$$
with the symmetric matrix
$$
g_{ij}(A)
\equiv \sum _{s=0}^i\sum _{r=0}^j {A_s^{(i)}A_r^{(j)}\over (i+j-s-r+1)}
=\int _0^1 {\rm d} x\; \left( \sum _{s=0}^iA_s^{(i)}x^{i-s}\right) \left(
\sum _{r=0}^j A_r^{(j)}x^{j-r}\right)
\eqno(91)
$$
Once more we have reduced the issue of diagonalizing 2-point functions to a
problem of orthogonal polynomials.
Notice the differences with respect to the purely bosonic case: the
integral goes from 0 to 1 now, and the polynomials have no definite
parity. The solution is given in terms of the $\pi$-polynomials defined as
$$
\pi_i(x) =  P_i (2x-1) = \sum _{s=0}^i A_s^{(i)} x^{i-s}\qquad ,\quad
A_s^{(i)} = {(-1)^s\over s!} {(2i - s)! \over [(i-s)!]^2}
\quad ,\eqno(92)
$$
for which $g_{ij}= {1\over 2i+1}\delta_{ij}$. The corresponding coefficients
$B_n$ are given by
$$
B_{m-s}={A_{s-1}^{(m-1)}-A_s^{(m)}\over 2m}\qquad ,\qquad B_m=-{A_0^{(m)}\over
2m} \quad .
\eqno(93)
$$
Also the choice (92) guarantees that the only non-vanishing 1-point functions
are
$$
\left\langle \widehat \sigma _{m-1}\right\rangle  = -{1\over 2\kappa ^2}
{\mu^{2m}\over 2m(2m-1)}\quad ,\qquad
\left\langle \widehat \sigma _{m}\right\rangle  = {1\over 2\kappa^2}
{\mu^{2m+1}\over 2m(2m+1)}\quad ,
\eqno(94)
$$
the former corresponds to the average ``area" to be compared with Eq. (71),
while the latter is expected to represent part of the average energy, as in the
bosonic case.

Therefore we propose the following definition of the superconformal basis of
scaling operators
$$\eqalign{
\widehat \sigma _n&=\sum _{s=0}^n {(-1)^s\over s!}{(2n-s)!\over \left[
(n-s)!\right] ^2}\mu ^s\sigma _{n-s} \quad ,\cr
\widehat \tau _n^\pm &=\sum _{s=0}^n {(-1)^s\over s!}{(2n-s)!\over \left[
(n-s)!\right] ^2}\mu ^s\tau _{n-s}^\pm \quad ,\cr}
\eqno(95)
$$
which is our main result. We emphasize, though, that Eq. (95)
corresponds to a lowest-order transformation. It can also be inverted, so that
the super-KdV frame reads
$$\eqalign{
\sigma _n&=\left[ n!\right] ^2\sum _{s=0}^n {(2n+1-2s)\over s!(2n+1-s)!}\mu
^s\widehat \sigma _{n-s}\quad ,\cr
\tau _n^\pm &=\left[ n!\right] ^2\sum _{s=0}^n {(2n+1-2s)\over s!(2n+1-s)!}\mu
^s\widehat \tau _{n-s}^\pm \quad .\cr }
\eqno(96)
$$
We also notice that, within the approximation (86)-(88), we have
$$
\widehat D_\pm =\sum _i\widehat \tau _i^\pm {\partial \over \partial \widehat
t_i} =\sum _i \tau _i^\pm {\partial \over \partial t_i}=D_\pm
\quad ,\eqno(97)
$$
and the free energy can still be written as ${\cal F}_s={1\over 2}(1-\widehat
D_+\widehat D_-){\cal F}_{bosonic}$.
In the new regime we verify that the free energy scales as
$$
{\cal F} \sim \mu^{m+1/2}\quad ,
\eqno(98)
$$
while general $N$-point functions scale as
$$
\left\langle \prod_i \widehat {\cal O}_i\right\rangle  \sim \mu^{m+1/2 + \sum
_i\delta_i}\quad ,
\eqno(99)
$$
with the scaling exponents
$\delta_{\sigma_n}  = n-m$ and $\delta_{\nu_n^\pm }  = n-m \mp {1\over 2}$, to
be compared with the continuum values $-\beta/\alpha _+$.

The $m$-th critical string equation becomes
$$
\sum _nB_n\mu ^{m-n}u^n={u^m\over 2m}\left[ \pi _{m-1}\!\left( {u\over \mu
}\right) - \pi _m\! \left( {u\over \mu }\right) \right] =0 \quad ,
\eqno(100)
$$
in agreement with the solution $u\to \mu $ due to the general property $\pi
_n(1)=P_n(1)=1$ of the $\pi $-polynomials.

\vskip .5truecm
\noindent 3.4  {\it Super-Liouville/eigenvalue dictionary}
\vskip .3truecm

\noindent
As in the bosonic case, we suppose that the $m$-th critical supersymmetric
model
corresponds to some dressed $(p,q)$-minimal superconformal theory; we determine
the numbers $p$ and $q$ by analysing the scalings in two different regimes,
namely the $t$- and $\mu $-averaged regimes.

In the continuum theory the partition function behaves as $Z\sim t^{-Q/\beta
_{min}}$ and $Z\sim \mu ^{-Q/\alpha _+}$. For the supersymmetric $(p,q)$
model, we have $\alpha _0={p-q\over 2\sqrt {pq}}$ and
$$
-{Q\over \beta _{min}}={2(p+q)\over (p+q)-2}\qquad ,\qquad -{Q\over \alpha
_+}=1+{q\over p}\quad .
\eqno(101)
$$
In the discrete theory we have found ${\cal F}_s \sim t^{2+1/m}$ and ${\cal
F}_s\sim \mu ^{2m+1}$, thus implying the following pair of equations:
$$\eqalign{
{2(p+q)\over (p+q)-2}&=2+{1\over m}\quad ,\cr
1+{q\over p}&=2m+1 \quad ,\cr }
\eqno(102)
$$
whose solution\ref{3} is $(p,q)=(2,4m)$, $m=1,2,\cdots $. The corresponding
central charges are $c={3\over 2}\left[ 1-{(2m-1)^2\over m}\right] =0,-{21\over
4}, \cdots $. The case $m=1, c=0$ corresponds to pure supergravity and is the
only unitary model within this series.

Concerning the spectrum of primary operators, we have two sectors (NS and R):
the total number of primaries is ${1\over 2}((p-1)(q-1)+1)$, which makes up
$2m$ operators ($m$ in each sector) in our case. The conformal weight $\Delta
_{rr'}$ of such operators is given by the formula
$$
\Delta _{rr'}={(rp-r'q)^2-(q-p)^2\over 8pq}+{1-(-1)^{r-r'}\over 32}\quad ,
\eqno(103)
$$
with $1\le r'\le p-1$ (i.e. $r'=1$ in our case) and $1\le r\le q-1=4m-1$. When
the difference $r-r'$ is even, the operator belongs to the NS sector,
$$
\Delta ^{{}^{NS}}_{2i+1,1}={i(i-2m+1)\over 4m}\qquad ,\quad i=0,1,\cdots
\, ,\, 2m-1\quad ,
\eqno(104)
$$
and for $r-r'$ odd one has the R sector,
$$
\Delta ^{{}^R}_{2i+2,1}={(i+1-m)^2-(m-1/2)^2\over 4m}+{1\over 16}\qquad ,\quad
i=0,1,\cdots \, ,\, 2m-2 \quad .
\eqno(105)
$$
In fact we can restrict the label $i$ to the range $i=0,\cdots ,m-1$, since the
remaining values only duplicate the spectrum.
{}From the weights $\Delta _{rr'}$ it is straightforward to calculate the
corresponding momenta $k_{rr'}$ and dressing charges $\beta _{rr'}$. Then, by
comparing the values of $-\beta _{rr'}/\alpha _+$ with the exponents in
Eq. (99) we can associate the bosonic scaling operators with the NS vertices,
$$
\widehat \sigma _n \; \leftrightarrow \; \psi _{_{NS}}(k_{_{2(m-n)-1,1}}) \quad
,\quad n=0,\cdots , m-1 \quad ,
\eqno(106a)
$$
and the fermionic operators with the R vertices in the following subtle manner
$$\eqalignno{
\widehat \nu _n^+ \; &\leftrightarrow \; V_{-{1\over 2}}(k_{_{2(m-n),1}})\quad
,&(106b)\cr
\widehat \nu _n^- \; &\leftrightarrow \; \mu ^2\psi
_{_{NS}}^{^{(-1)}}(0)V_{-{1\over 2}}(2\alpha _0-k_{_{2(m-n),1}})\quad ,\quad
n=0,\cdots ,m-1\quad .
&(106c)\cr}
$$
The special operator $\widehat \sigma _{m-1}$ corresponds to the ``area" or
cosmological term. We can also define an energy operator,
$$
\varepsilon _m=\widehat \sigma _m-{(2m-1)\over (2m+1)}\mu \widehat \sigma
_{m-1} \quad ,
\eqno(107)
$$
which points out the role of the operator $\widehat \sigma _m$. The remaining
operators correspond to secondary fields. Notice the absence of a boundary
operator, as opposed to the bosonic case.

\vskip .7truecm
\noindent 3.5 {\it Wave functions and minisuperspace approximation}
\vskip .3truecm

\noindent
We shall determine the supersymmetric version of Eq. (15) obeyed by the wave
functions of scaling operators in the superconformal frame. In this way we
expect to circumvent some ambiguities, which usually arise in
attempts\ref{16} to derive the super WdW equation in minisuperspace
approximations from the continuum theory.

We start from the macroscopic superloop $W_\pm (l,\theta _\pm)$ defined in
Ref.  [3], whose expectation value reads
$$
\left\langle  W_\pm (l,\theta_\pm )\right\rangle ={1\over
\kappa \sqrt {\pi l}}\left( -\partial_t^{-1} +\tau_+\tau_-\partial_t \pm \theta
_\pm\partial_t^{-1}\tau_\mp\partial _t\right) \E^{-lu}\quad .
\eqno(108)
$$
Using the string equation and the definition (78$b$), this equation can be
rewritten as
$$
\langle W_\pm (l,\theta _\pm)\rangle =\left[ 1-D_+D_-\mp
\theta _\pm D_\mp \right]
{1\over \kappa \sqrt {\pi l}}\int _{t_0}^\infty \!{\rm d}y\, \E ^{-lu(y)}
\quad ,\eqno(109)
$$
to be compared with Eq. (23). In fact we have the same integral of the bosonic
model, implying an expansion in terms of the functions $\psi _n$ defined in
Eq. (25). Using the following identity between Bessel functions and
$\pi$-polynomials,
$$
\E ^{-x/2}K_{n+1/2}\left( {x\over 2}\right) = \sqrt {\pi \over x} \int
_x^\infty \! {\rm d}z\, \pi _n\! \left( {z\over x}\right) \E ^{-z} \quad ,
\eqno(110)
$$
we can calculate the integral in Eq. (109) in the superconformal frame, finding
as a result the following expansion
$$\eqalignno{
{1\over \kappa \sqrt {\pi l}}\int _{t_0}^\infty \!{\rm d}y\, \E ^{-lu(y)}&=
{1\over \kappa \pi }{u^{m+1/2}\over 2m}\E ^{-lu/2} \left[ K_{m-1/2}\!\left(
{lu\over 2}\right) -K_{m+1/2}\!\left( {lu\over 2}\right) \right]\cr
&+{1\over \kappa \pi }\sum _{n=0}^\infty \widehat t_nu^{n+1/2}\E
^{-lu/2}K_{n+1/2}\!\left( {lu\over 2}\right) \quad .&(111)\cr }
$$
We therefore obtain, in the regime $\widehat t_n, \widehat \tau _n^\pm \to 0$,
the following wave functions
$$\eqalign{
\Psi _i^{^{NS}}&=\langle \widehat \sigma _iW_\pm (l,\theta _\pm )\rangle =
{1\over \kappa \pi }\mu ^{i+1/2}\E ^{-l\mu /2}K_{i+1/2}\!\left( {l\mu \over
2}\right) \quad ,\cr
\Psi _{i\pm}^{^R}&=\langle \widehat \nu _i^\pm W_\pm (l,\theta _\pm)\rangle
=\mp \theta _\mp \Psi _i^{^{NS}}\quad .\cr }
\eqno(112)
$$
Both satisfy the wave equation
$$
\left\{ -\left( l{\partial \over \partial l} + {\mu l\over 2} \right) ^2
+ {\mu^2 l^2 \over 4} + \left( i+{1\over 2}\right) ^2\right\} \Psi _i =0
\; ,\eqno(113)
$$
which we expect to be the supersymmetric WdW equation in the minisuperspace
approximation. Indeed, after dropping the space dependence of the
fields in the continuum theory, we have the super-Liouville Hamiltonian
$$
H= {1\over 2} p^2 + {\mu\over 2\alpha_+^2}\overline \psi\psi \E^{{1\over 2}
\alpha_+\phi}+ {\mu^2\over 2\alpha_+^2} \E^{\alpha_+\phi} + {Q^2\over 8}
\quad ,\eqno(114)$$
suggesting\ref{17} the associations
$$\eqalignno{
p^2 + {\mu\over \alpha_+^2} \overline \psi\psi \E^{{1\over 2}\alpha_+\phi}&\to
-\left( l{\partial\over\partial l} + {\mu l\over 2} \right)^2\quad ,&(115a)\cr
{\mu^2 \over 2\alpha_+^2} \E^{\alpha_+\phi}  & \to \mu^2 l^2\quad .&(115b)\cr}
$$
Notice that the fermionic contribution is summarized by an $l$-dependent
shift of the momentum $p$, which amounts to the extra exponential factor $\E
^{-l\mu /2}$ in comparison to the Bessel wave functions of the bosonic model.
The understanding of this effect in terms of the continuum formulation is at
present under investigation.

\vskip .4truecm
\noindent 4. {\bf Conclusion}
\vskip .3truecm

\noindent Even though a matrix formulation is still missing, we have found the
superconformal background of couplings and operators of the discrete
super-eigenvalue model, which can be completely translated into the
super-Liouville language, and a precise correspondence to the dressed $(2,4m)$
minimal superconformal theory can be established. As a by-product we have
derived the wave equation that should be equivalent to the proper
minisuperspace
approximation of the WdW equation that supersymmetric loop amplitudes are
expected to obey.

\vskip .4truecm
\noindent {\bf References}
\vskip.3truecm

\refer[[1]/G. Moore, N. Seiberg and M. Staudacher, Nucl. Phys. {\bf B362}
(1991) 665.]

\refer[[2]/E. Abdalla, M.C.B. Abdalla, D. Dalmazi and  K. Harada, Phys. Rev.
Lett. {\bf 68} (1992) 1641, Int. J. Mod. Phys. {\bf A7} (1992) 2437, IFT-Prep.
042/91;]

\refer [/L. Alvarez-Gaum\'e and  Ph. Zaugg, Phys. Lett. {\bf B273} (1991) 81;]

\refer[/K. Aoki and E. D'Hoker, Mod. Phys. Lett. {\bf A7} (1992) 333.]

\refer[[3]/L. Alvarez-Gaum\'e, H. Itoyama, J.L. Ma\~nez and A. Zadra, Int. J.
Mod. Phys. {\bf A7} (1992) 5337.]

\refer[[4]/J. Distler and H. Kawai, Nucl. Phys.  {\bf B321 }(1988) 171;]

\refer[/F. David,  Mod. Phys. Lett. {\bf A3 }(1988) 1651.]

\refer[[5]/N. Seiberg, Lecture at 1990 Yukawa Int. Sem. on Common Trends in
Math.
and Quantum Field Theory, and Carg\`ese Meeting on Random Surfaces, Quantum
Gravity and Strings, 1990.]

\refer[[6]/M. Goulian and  M. Li, Phys. Rev. Lett. {\bf 66} (1991) 2051.]

\refer[[7]/P. di Francesco and  D. Kutasov, Phys. Lett. {\bf 261B} (1991) 385;
Nucl. Phys. {\bf B375} (1992) 119.]

\refer[[8]/V.A. Kazakov, Mod. Phys. Lett. {\bf A4} (1989) 2125.]

\refer[[9]/D.J. Gross and A.A.  Migdal, Nucl. Phys. {\bf B340} (1990) 333;]

\refer[/M.R. Douglas and S.H. Shenker, Nucl. Phys. {\bf B335} (1990) 635;]

\refer[/E. Br\'ezin and V.A. Kazakov, Phys. Lett. {\bf 236B} (1990) 144.]

\refer[[10]/E. Abdalla, M.C.B. Abdalla, D. Dalmazi and A. Zadra, {\it 2D
Gravity in Non-Critical Strings: continuum and discrete approaches}, Lecture
Notes in Physics, series M, Springer Verlag (to appear).]

\refer[[11]/J. Distler, Z. Hlousek and H. Kawai, Int. J. Mod. Phys. {\bf A5}
(1990) 391.]

\refer[[12]/M. Green and N. Seiberg Nucl. Phys. {\bf B299} (1988) 559.]

\refer[[13]/D. Friedan, E. Martinec and S. Shenker, Nucl. Phys. {\bf B241}
(1986) 93.]

\refer[[14]/L. Alvarez-Gaum\'e, K. Becker, M. Becker,
R. Empar\'an and J.L. Ma\~nes, Int. J. Mod. Phys. {\bf A8} (1993) 2297.]

\refer[[15]/K. Becker and M. Becker,  Mod. Phys. Lett. {\bf A8} (1993) 1205.]

\refer[[16]/P.D. D'Earth and D.I. Hughes, Nucl. Phys. {\bf B378} (1992) 381.]

\refer[[17]/E. Abdalla and A. Zadra, Trieste Summer School and Workshop in High
Energy Physics and Cosmology, 1993, to appear]

\end